\documentclass[]{emulateapj}

\newcommand{\msol}{M_\odot}
\def\rhiz{$\rho_{\rm HI}(z)$}

\def\rhi{$\rho_{\rm HI}$}
\def\rhodot{$\dot{\rho_{*}}$}
\def\rhodotz{$\dot{\rho_{*}}$$(z)$}
\def\mrhi{\rho_{\rm HI}}

\newcommand{\nhi}{$N_{\rm HI}$}
\newcommand{\mnhi}{N_{\rm HI}}

\newcommand{\fnhi}{$f(\mnhi,X)$}
\newcommand{\lox}{$\ell(X)$}
\newcommand{\mfnhi}{f(N_{\rm HI},X)}

\newcommand{\ndla}{738}

\newcommand{\cm}[1]{\, {\rm cm^{#1}}}
\newcommand{\mkms}{{\rm \; km\;s^{-1}}}

\newcommand{\lya}{Ly$\alpha$}

\begin{document}

\title{On the (Non)Evolution of \ion{H}{1} Disks over Cosmic Time}

\author{J. Xavier Prochaska\altaffilmark{1} and
	Arthur M. Wolfe\altaffilmark{2}
}

\altaffiltext{1}{Department of Astronomy and Astrophysics, 
UCO/Lick Observatory;
University of California, 1156 High Street, Santa Cruz, CA  95064;
xavier@ucolick.org}
\altaffiltext{2}{Department of Physics, and 
Center for Astrophysics and Space Sciences, 
University of California, San
Diego, 
Gilman Dr., La Jolla; CA 92093-0424; awolfe@ucsd.edu}

\begin{abstract}
We present new results on the frequency distribution of projected
\ion{H}{1} column densities \fnhi, total comoving
covering fraction,
and integrated mass densities \rhi\ of high redshift, \ion{H}{1}
`disks' from a survey of damped \lya\ systems (DLAs) in the
Sloan Digital Sky Survey, Data Release 5.
For the full sample spanning $z=2.2$ to 5 (738 DLAs), \fnhi\
is well fitted by a double power-law with a break column density
$N_d = 10^{21.55 \pm 0.04} \cm{-2}$ and low/high-end exponents
$\alpha = -2.00 \pm 0.05, -6.4^{+1.1}_{-1.6}$.
The shape of \fnhi\ is invariant during this redshift interval 
and also follows the projected surface density 
distribution of present-day \ion{H}{1} disks as inferred from
21cm observations.   We conclude that \ion{H}{1} gas has been
distributed in a self-similar fashion for the past 12\,Gyr.
The normalization of \fnhi, in contrast,
decreases by a factor of two during the $\approx 2$\,Gyr interval
from $z=4$ to 2.2 giving corresponding decreases in both the 
total covering fraction and \rhi.
At $z \approx 2$, these quantities match the present-day values
suggesting no evolution during the past $\approx 10$\,Gyr.
We argue that the evolution at early times is driven by `violent'
processes that removes gas from nearly half the
galaxies at $z \approx 3$ establishing the antecedants of 
current early-type galaxies.  The perceived
constancy of \rhi, meanwhile, implies that 
\ion{H}{1} gas is a necessary but insufficient pre-condition
for star formation and that the
global star-formation rate is driven 
by the accretion and condensation of fresh gas from the intergalactic medium.
\end{abstract}


\section{Introduction}

In the current paradigm of galaxy formation within CDM cosmology,
baryons accrete, dissipate, and settle to the
centers of dark matter halos.  
Gas with a non-negligible angular momentum will
form an \ion{H}{1} `disk' with typical surface densities
exceeding $1 \msol \, \rm pc^{-2}$ 
(or \ion{H}{1} column densities, $\mnhi > 10^{20} \cm{-2}$).
Various processes (e.g.\ merger induced
shocks, secular evolution) inspire the formation of molecular clouds
that cool, fragment, and initate star formation.  
Finally, stellar feedback
(e.g.\ winds, supernovae), AGN activity, galaxy interactions,
and even magnetic fields \citep{wolfe08} 
may inhibit star formation, 
perhaps driving the gas from the galaxy. 

The \ion{H}{1} disks of galaxies, therefore, serve as a 
barometer of recent star formation activity and a record of prior processing. 
The mass, metallicity, velocity field, surface density profile, etc.\
reflect both the underlying dark matter potential and also the
star formation history of the galaxy.  In the local universe,
\ion{H}{1} disks are mapped in the 21cm line with radio telescopes.
These data reveal the mass, surface density profiles, and kinematics of
modern \ion{H}{1} disks \citep[e.g.][]{zms+05,things08}.
With current facilities, unfortunately, it is impossible to survey
\ion{H}{1} disks in 21cm emission at high redshift.
Such analysis awaits the construction of facilities like the proposed 
Square Kilometer Array.

In lieu of 21cm observations, one may observe \ion{H}{1} gas through
electronic transitions, e.g. the Lyman, Balmer, and Paschen series. 
Although these lines can be studied in emission, they
arise in ionized gas via recombination
processes.  Furthermore, detectable 
fluxes require a strong (i.e.\ local) excitation/ionization source 
and one tends to map special, isolated regions of the galaxy.
To study the bulk of an \ion{H}{1} disk,
one may instead probe the gas in absorption 
via the Lyman series\footnote{Although 21cm absorption also traces
\ion{H}{1} gas, the optical depth is inversely proportional to temperature
and one is primarily sensitive to cold, \ion{H}{1} gas.}  \citep{wolfe86}.  
At the characteristic column densities of \ion{H}{1} disks, 
the \lya\ transition is damped
and astronomers refer to the observed profiles as damped \lya\ systems
\citep[DLAs;][]{wgp05}.  These DLA profiles are mainly
revealed in the spectra of distant quasars, yet they also manifest in the
spectra of GRB afterglows \citep[e.g.][]{cpb+05}. 
Unfortunately, these intrinsically luminous sources cover only a small 
fraction of the sky such that one rarely intesects a given \ion{H}{1}
disk with multiple sightlines \citep[e.g.][]{ehm+07}.
Therefore, \ion{H}{1} disks at high $z$ must be
studied statistically through the observation
of thousands of quasars across the sky.  

\begin{figure*}
\begin{center}
\includegraphics[height=6.8in,angle=90]{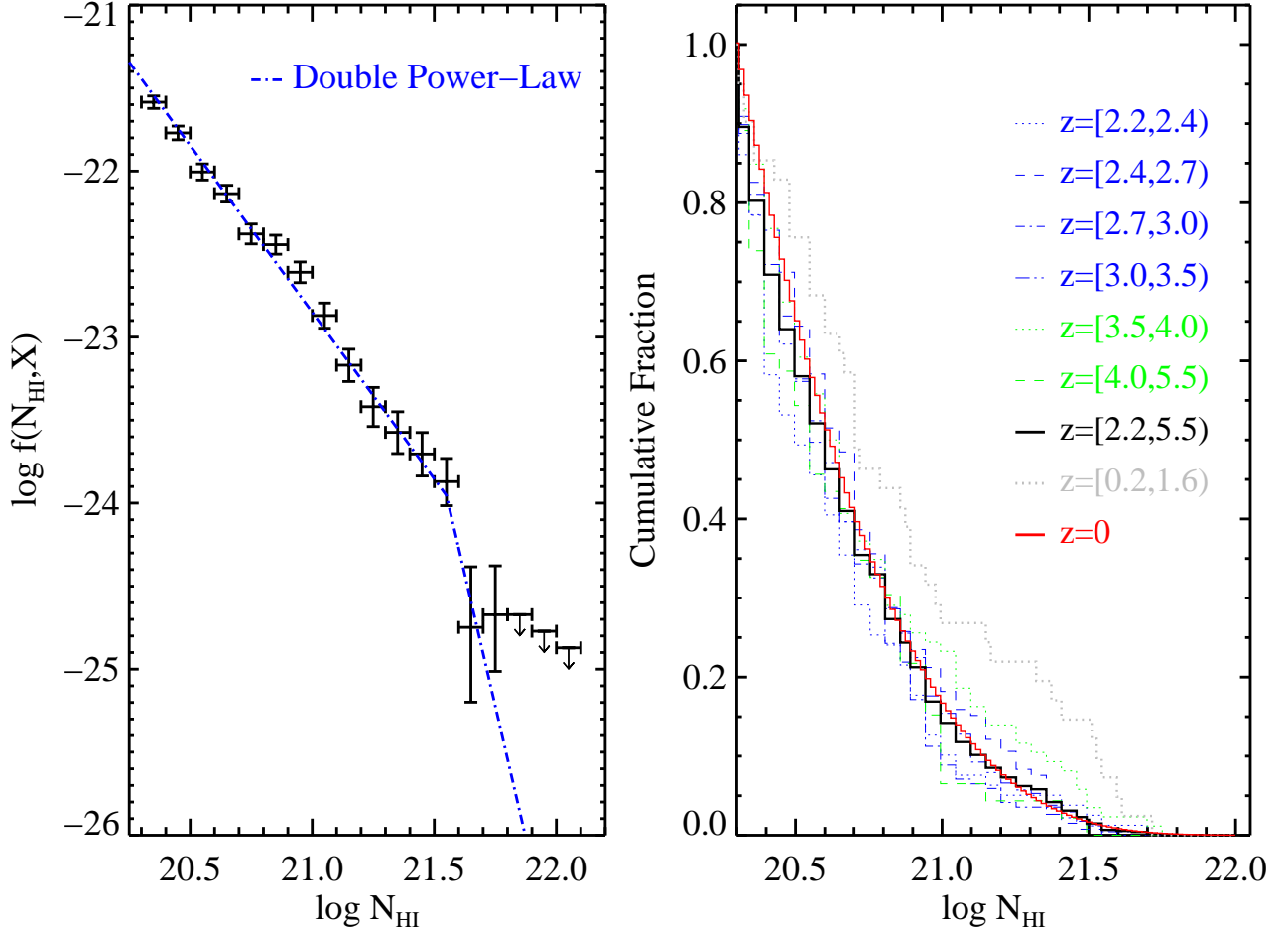}
\end{center}
\caption{
{\it Left:}  The integrated frequency distribution \fnhi\ of 
projected \ion{H}{1} column densities for galaxies
at $z = 2$ to 4.  The overplotted curve represents
the best-fit, double power-law which has a break column density
$N_d = 10^{21.55} \cm{-2}$, a `faint-end' exponent $\alpha_3 = -2.00 \pm 0.05$,
and a high-end exponent $\alpha_4 < -4.4$ (95\%\ c.l.).
({\it right}) Cumulative distribution functions of \fnhi\
for galaxies in a series of redshift intervals.  For $z>2$, the
shape of \fnhi\ is invariant and, remarkably, matches
with the observed function for local \ion{H}{1} disks
\citep{zvb+05}.  The gray dotted line shows the results
for $z \sim 1$ from \cite{rtn06}.
}
\label{fig:fnhi}
\end{figure*}

This experiment has been realized over the past
few years as an unintended consequence of the Sloan Digital
Sky Survey (SDSS) of high redshift quasars 
\citep[][; hereafter PHW05]{ph04,phw05}.
In this Letter, we report on the results for a survey of the
SDSS Data Release 5 \citep[SDSS-DR5;][]{sdssdr5}.  We place new
constraints on the projected \ion{H}{1} column density 
distribution, total covering fraction, and integrated mass density
of \ion{H}{1} disks at $z>2$.
We search for evolution in these quantities from $z=2$ to 4
and also compare the measurements with \ion{H}{1} disks from the
local universe \citep{zvb+05}.  These results offer new insight on the evolution
of \ion{H}{1} disks and their role in the processes of galaxy formation.
Throughout the Letter, we adopt a $\Lambda$CDM cosmology
with $\Omega_\Lambda = 0.7$, $\Omega_m = 0.3$, 
and $H_0 = 72 \mkms \rm Mpc^{-1}$.


\section{Results}

We have surveyed damped \lya\ systems at $z \ge 2.2$ using the 
database of quasar spectroscopy from the SDSS-DR5.
We have implemented the algorithms developed in PHW05 to search for DLA
candidates, measure the survey path, eliminate strong BAL quasars,
and to fit Voigt functions to candidate DLAs.
Tables and figures for the \lya\ fits are provided online
(http://www.ucolick.org/$\sim$xavier/SDSS).
The full statistical sample now comprises \ndla\ DLAs with 
$\mnhi \ge 10^{20.3} \cm{-2}$ over a total redshift path
$\Delta z = 3082.5$, each located 
at velocity $\delta v > 3000\mkms$
from the background quasar.

In Figure~\ref{fig:fnhi}a we present the \nhi\ frequency distribution
\fnhi\ of the full statistical sample.  
This measure describes the projected column density
distribution of \ion{H}{1} gas in galaxies at high $z$ per comoving
absorption pathlength $dX$.  Similar to our previous results 
(PHW05), \fnhi\ is well described by a power-law at low \nhi\ 
values $\mfnhi \sim \mnhi^\alpha$ with $\alpha \approx -2$, but
transitions to a steeper function at $\mnhi \approx 10^{21.5} \cm{-2}$.
This break in \fnhi\ is required to yield a finite \ion{H}{1} mass density,
$\mrhi = (m_p H_0 / c) \int N \mfnhi dX dN$.
Following the formalism in PHW05, we fitted a double power-law to
\fnhi;  the best-fit model is overplotted on the data in Figure~\ref{fig:fnhi}a
and tabulated in Table~\ref{tab:newsumm} for a series of redshift intervals.
The table also lists the zeroth and first moments of \fnhi\
which give the line density \lox\ and \rhi\ values respectively.
The former quantity represents the covering fraction 
per $dX$ for the integrated population of \ion{H}{1} disks at a given epoch.
The latter quantity is the comoving mass density of \ion{H}{1}
gas in high $z$ galaxies.

\begin{deluxetable*}{lccccccccccc}
\tablewidth{0pc}
\tablecaption{DR5 SUMMARY\label{tab:newsumm}}
\tabletypesize{\footnotesize}
\tablehead{\colhead{z} & \colhead{$\Delta X$} & \colhead{$\Delta z$} & \colhead{$m_{\rm DLA}$} & \colhead{$\bar z^a$} & 
\colhead{$k_3$} & \colhead{$\log(N_d/\cm{-2})$} & \colhead{$\alpha_3$}& \colhead{$\alpha_4$} & 
\colhead{\lox$^b$} & \colhead{\rhi$^c$}\\
& & & & & & & & & & ($10^8 \msol \, \rm Mpc^{-3}$) }
\startdata
$\lbrack$2.2,5.5]&10872.8& 3082.5& 739&3.05&$-23.95^{+0.02}_{-0.02}$&$21.55^{+0.04}_{-0.03}$&$-2.01^{+0.05}_{-0.05}$&$ -6.34^{+1.06}_{-1.60}$&$0.068^{+0.003}_{-0.003}$&$0.851^{+0.046}_{-0.045}$\\
$\lbrack$2.2,2.4]& 1652.7&  514.4&  79&2.31&$-24.68^{+0.05}_{-0.05}$&$21.70^{+0.12}_{-0.07}$&$-2.27^{+0.16}_{-0.18}$&$-10.00^{+4.69}_{-0.00}$&$0.048^{+0.006}_{-0.005}$&$0.555^{+0.095}_{-0.096}$\\
$\lbrack$2.4,2.7]& 2405.8&  717.5& 132&2.57&$-23.54^{+0.04}_{-0.04}$&$21.40^{+0.07}_{-0.03}$&$-1.73^{+0.12}_{-0.13}$&$ -7.18^{+1.79}_{-1.79}$&$0.055^{+0.005}_{-0.005}$&$0.736^{+0.083}_{-0.087}$\\
$\lbrack$2.7,3.0]& 2539.7&  723.7& 169&2.86&$-24.17^{+0.03}_{-0.03}$&$21.60^{+0.10}_{-0.06}$&$-2.12^{+0.11}_{-0.12}$&$-10.00^{+3.39}_{-0.00}$&$0.067^{+0.006}_{-0.005}$&$0.743^{+0.077}_{-0.078}$\\
$\lbrack$3.0,3.5]& 2702.5&  732.2& 227&3.22&$-23.85^{+0.03}_{-0.03}$&$21.55^{+0.09}_{-0.05}$&$-2.00^{+0.09}_{-0.10}$&$ -8.62^{+2.56}_{-0.93}$&$0.084^{+0.006}_{-0.006}$&$1.029^{+0.086}_{-0.090}$\\
$\lbrack$3.5,4.0]& 1139.2&  291.3&  86&3.70&$-24.16^{+0.05}_{-0.05}$&$21.75^{+0.08}_{-0.06}$&$-1.86^{+0.13}_{-0.14}$&$-10.00^{+3.85}_{-0.00}$&$0.075^{+0.009}_{-0.008}$&$1.213^{+0.210}_{-0.211}$\\
$\lbrack$4.0,5.5]&  432.8&  103.6&  46&4.39&$-23.76^{+0.07}_{-0.07}$&$21.50^{+0.22}_{-0.07}$&$-2.13^{+0.21}_{-0.24}$&$-10.00^{+4.43}_{-0.00}$&$0.106^{+0.018}_{-0.016}$&$1.179^{+0.251}_{-0.222}$\\
\enddata
\tablenotetext{a}{Mean absorption redshift of the DLA sample.}
\tablenotetext{b}{Line density of DLAs per absorption length $dX$.  Also written as $dN/dX$ in the literature.}
\tablenotetext{c}{Note that we recover the same estimate for \rhi\ whether we sum the discrete \nhi\ values or integrate the best fitting double power-law (see also PHW05).}
\end{deluxetable*}

A principal result of our survey is that the
shape of the \ion{H}{1} distribution function does not evolve.
Figure~\ref{fig:fnhi}b shows the cumulative
\fnhi\ distributions for a series of redshift intervals from $z=2.2$ to 5.
We have performed a series of two-sided Kolmogorov-Smirnov test 
comparisons and find that the null hypothesis cannot be ruled 
out at greater than 90\%\ confidence for any pair.
Table~\ref{tab:newsumm} also reveals that the parameters of a 
double power-law fit to the data show no significant ($>2\sigma$)
variations with redshift.

In Figure~\ref{fig:fnhi}b, we also present the cumulative 
\fnhi\ function for \ion{H}{1} disks in the local universe, 
as estimated from 21cm observations \citep{zvb+05}.
Remarkably, the $z\sim 0$ distribution function is a near perfect
match to the high $z$ universe. This is a stunning result
noted first and independently by \cite{zvb+05} and PHW05.
Although the universe and the galaxies within
it have evolved substantially over the $\approx 10$\,Gyr 
interval from $z=3$ to today, the combined distribution of
\ion{H}{1} surface densities is invariant.
Stated differently, the population of galaxies at any epoch
shows a self-similar projected \ion{H}{1} surface density distribution.

\begin{figure}
\begin{center}
\includegraphics[width=3.5in]{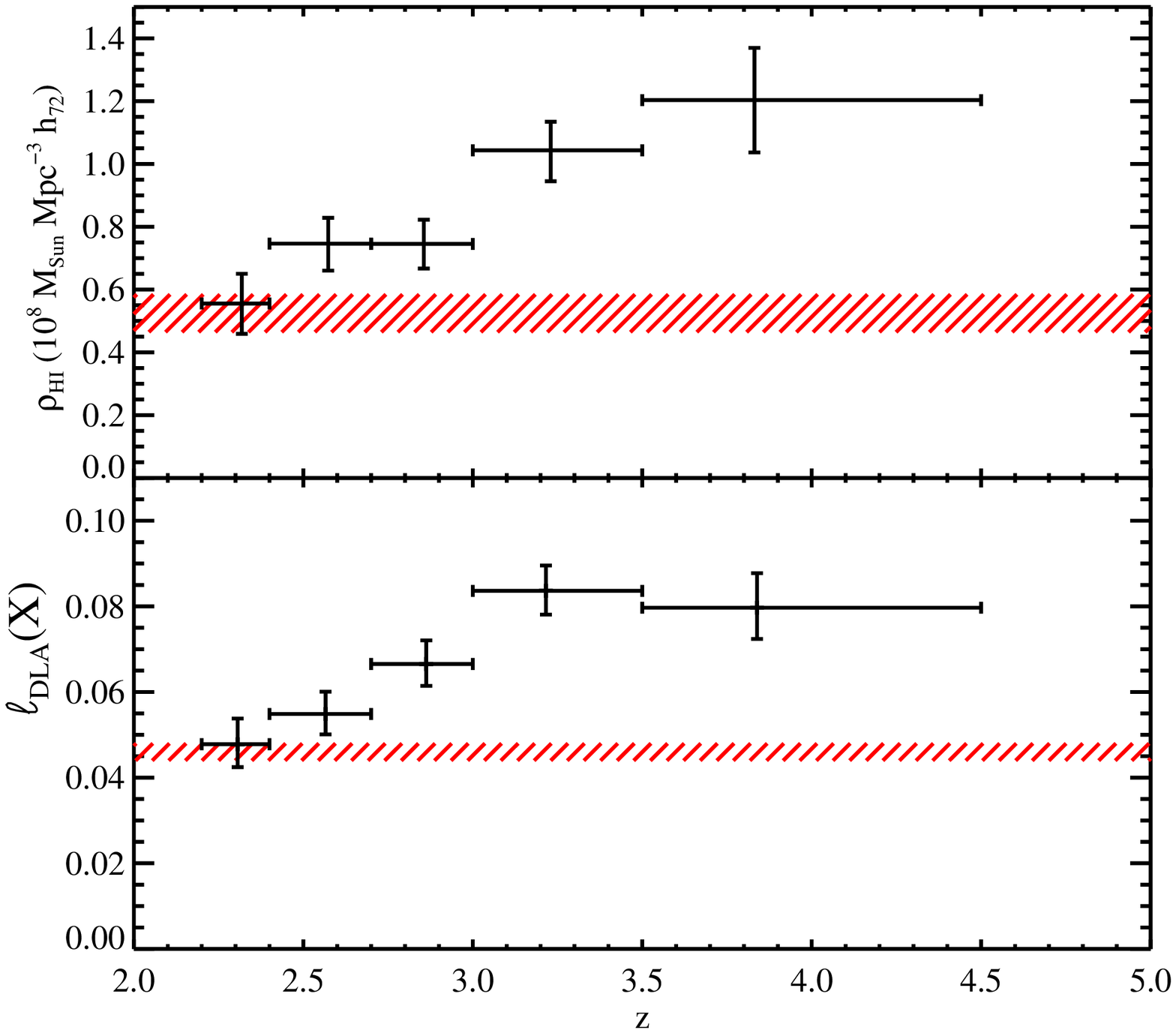}
\end{center}
\caption{
{\it Upper:}
Comoving \ion{H}{1} mass density \rhi\ of galaxies at $z>2$ assuming
a $\Lambda$CDM cosmology.  The \ion{H}{1} mass density is observed
to decline by $\approx 50\%$ from $z=4$ to 2.2, an interval
spanning less than 2\,Gyr.  The red band shows the estimate of \rhi\
at $z \sim 0$ from 21cm surveys of local \ion{H}{1} disks
\citep{zms+05}.  
{\it Lower:}
The line density of DLAs per comoving absorption length $dX$.
This quantity can be visualized as the integrated covering fraction
per comoving pathlength for \ion{H}{1} disks.
Folowing the mass density \rhi, the covering fraction decreases
by 50\%\ from $z \approx 4$ to 2 where it reaches the present-day value
(red band) as estimated from 21cm observations \citep{rws03,zms+05}.
Taken together, the results argue that \ion{H}{1} disks have not
evolved significantly over the past $\approx 10$\,Gyr.
}
\label{fig:moments}
\end{figure}

Although the shape of \fnhi\ is invariant, 
its normalization decreases with time.
This is revealed in Figure~\ref{fig:moments} where we present
\lox\ and \rhi\ from $z=5$ to 2.
Both the co-moving
covering fraction and the mass density of \ion{H}{1} disks
decreases by 50$\%$ in this $\approx 2$\,Gyr interval.
This sharp decline in both \lox\ and \rhi\ is a surprising
and profound result.  Before discussing its origin, we emphasize
that the evolution must occur at all column densities of the DLAs
such that the shape of \fnhi\ remains invariant.
Therefore, one should focus on processes that affect the 
inner and outer regions of \ion{H}{1} disks together.

\section{Discussion}

The results of the previous section have far reaching implications
for the nature and role of \ion{H}{1} disks and for the processes
of galaxy formation. 
Let us begin with the invariance in the shape of \fnhi.
In terms of statistical power,
the shape of \fnhi\ is dominated by systems with low \nhi\ values
and the primary result is that the `faint-end slope' of 
\fnhi\ is invariant.  In the local universe, 
low \nhi\ sightlines correspond to the outer regions of \ion{H}{1} disks.
We draw the inference that galaxies have self-similar 
surface density profiles in the outer disk at all cosmic times.

In the most straightforward, analytic models 
of galaxy formation \citep[e.g.][]{mmw98},
{\fnhi} is determined by the radial \ion{H}{1} surface density profiles, which
in turn are set by the total mass of
the system,  the angular momentum distribution
of the galaxy, the gas mass fraction, etc. 
This simple picture is modified by spiral density waves,
warps, galaxy mergers, the detailed nature of ISM clumping,
molecular cloud formation, and feedback from supernovae and/or
AGN activity.  All of these processes are expected to vary 
with time, especially the characteristic mass of galaxies.
The results presented in Figure~\ref{fig:fnhi}b suggest that the
outer regions of \ion{H}{1} disks are
not especially sensitive to these processes nor to the underlying
dark matter halo mass.  We note that this is actually a prediction
of viscous models of galactic disk formation \citep{lp87,obp91}. 
We await explorations of this topic witin the context of cosmological
simulations of galaxy formation \citep[e.g.][]{nsh04,rnp+06,pgp+08}.

Secular and feedback processes may be expected to 
have greater effect on the gas toward the inner regions, i.e.\ at
the highest surface densities. 
We have also searched for variations in \fnhi\ at large {\nhi},
but identify none. At the 95$\%$ c.l., all of the redshift
intervals have \fnhi\ distributions consistent with a break column
density of $N_d = 10^{21.6} \cm{-2}$.
Furthermore, restricting the frequency distributions 
to $\mnhi > 10^{21} \cm{-2}$, all give satisfactory KS-test probabilities.
We conclude there is no evolution at these column densities, 
but caution that the full sample includes only 105~DLAs.
The data also reveal, for the first time, that \fnhi\ is steeper than
$\alpha = -3$ beyond the break.  This cannot be attributed to projection
alone \citep[which predictes $\alpha = -3$; see][]{wc06},
but we associate the steeper drop
to the conversion of atomic gas to molecules \citep{schaye01_H2,zp06}.

\begin{figure*}
\epsscale{0.8}
\begin{center}
\includegraphics[height=6.8in,angle=90]{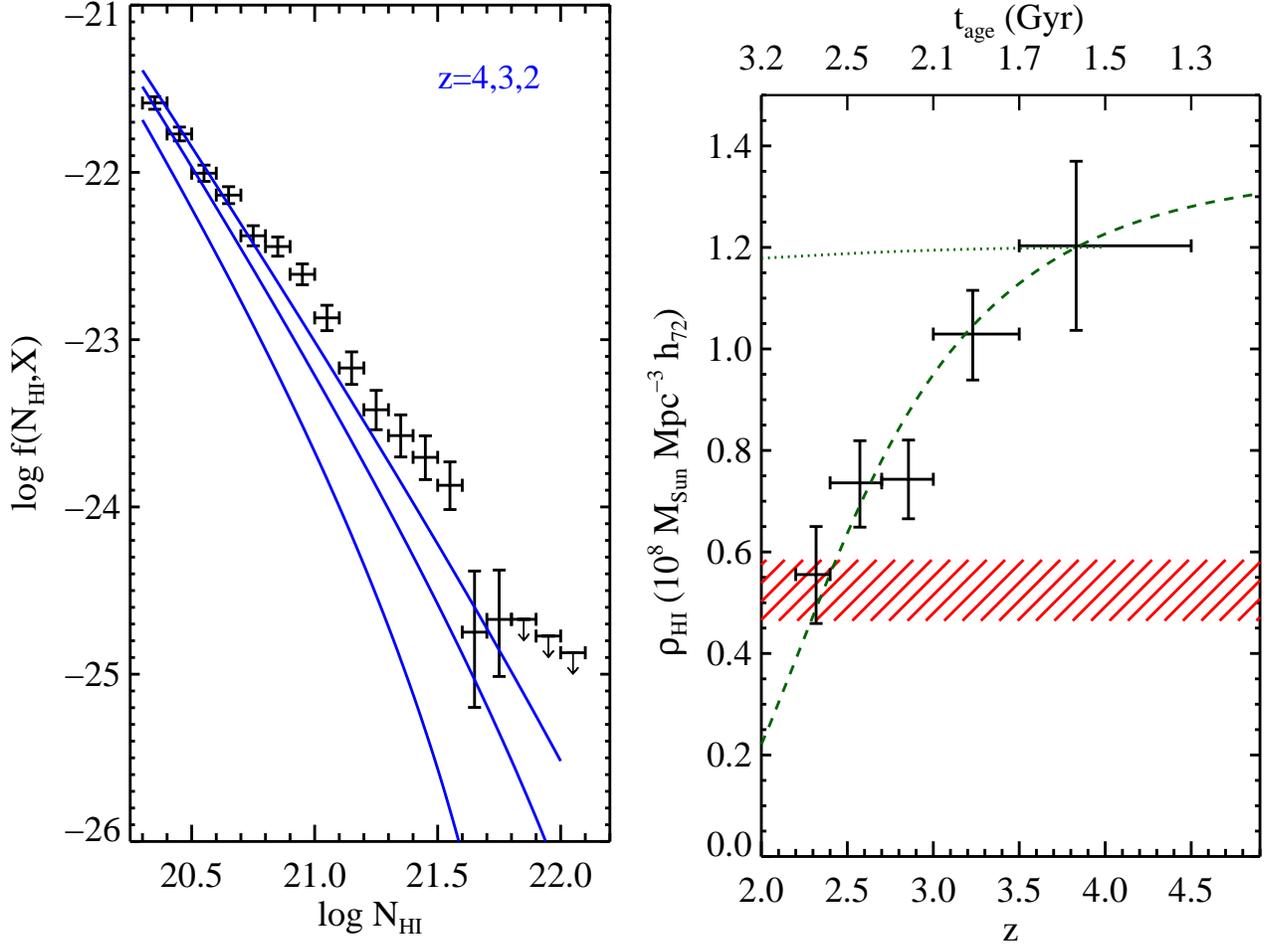}
\end{center}
\caption{
Models for the evolution of \fnhi\ and \rhi\ at high $z$. 
The blue solid curves in the left panel show the evolution in
\fnhi\ from $z=4$ to 2 preidcted by assuming star
formation occurs {\it in situ} according to the Schmidt law
($\Sigma_{SFR} \propto \Sigma_{\rm HI}^{1.4}$) in a closed
box model.  This model predicts too much evolution in \fnhi.
The smooth curves in the 
right-hand panel show the 
evoultion in \rhi\ predicted by the simple assumption that the
\ion{H}{1} consumption matches the mass of stars
formed from $z=5$ to 2 by integrating assumed star-formation histories.
The green dotted curve shows the results for the{\it in situ}
star-formation history of DLAs consistent with the upper limits
on \rhodotz\ set for the Hubble Ultra Deep Field (\cite{wc06}).
The green dashed curve shows the results for the star 
formation history of UV-selected galaxies
(\cite{bif+08}).  This curve is arbitrarily normalized to the 
observed \rhi\ value at $z=4$.
}
\label{fig:models}
\end{figure*}

Now consider the sharp decrease in the total comoving
covering fraction and
\ion{H}{1} mass density from $z=4$ to 2 (Figure~\ref{fig:moments},
Table~\ref{tab:newsumm}).
One's initial reaction may be to interpret this decline
in terms of active star-formation, i.e.\ the conversion 
of the \ion{H}{1} gas in DLAs to stars via {\it in situ} star formation.  
This interpretation is problematic for several reasons.
First, one expects star formation to mainly influence gas at
high \ion{H}{1} surface densities;  this is revealed, in part,
by the form of the Schmidt law,
$\Sigma_{\rm SFR}=K{\times}{\Sigma^{1.4}}$, where $\Sigma_{\rm SFR}$
is the SFR per unit area and $\Sigma$ is mass surface
density \citep{schmidt59,kennicutt98}. But the invariant
shape of {\fnhi} suggests that {\it in situ} star
formation in DLAs is unlikely to consume gas
according to the Schmidt law.
This is illustrated in Fig.~\ref{fig:models}a, which shows how an initial 
single power-law  approximation for {\fnhi} steepens with time
if  stars form according to the Schmidt law \citep{lwt95}. 
We find that the absence of changes in the shape of {\fnhi} implies that the
star-formation efficiency is less than 1/10 
that in local galaxies. 

Second, \cite{wc06} used the
infrequent detection of  extended, low surface-brightness galaxies
in the Hubble Ultra Deep field 
to set an upper limit on the comoving SFR density of
SFR density, {\rhodot}~$< 10^{-2.7} \msol \, \rm yr^{-1} \, Mpc^{-3}$. 
We used this limit to set an upper limit
on the decrease in {$\rho_{\rm HI}$} due to {\it in situ}
star formation. We fitted a conservative expression for {\rhodotz}
to be consistent with (a) this upper limit at z $\sim$ 3 and (b)
observations of star-forming galaxies
in the redshift interval $z$=[0,8] \citep[e.g.][]{bif+08}. 
We then computed the decrease in $\rho_{\rm H I}(z)$ by integrating {\rhodotz} from
$z$=6 to $z$. The resulting dip in $\rho_{\rm H I}$ is 
shown as the dotted curve in Fig.~\ref{fig:models}b. 
Clearly the decrease in $\rho_{\rm HI}$ predicted by {\it in situ}
star formation is too small to account for the factor of
two decrease observed. Specifically,
the total mass density of
stars formed in-situ in \ion{H}{1} disks is less than a few percent
of the observed \rhi\ at $z=4$.

Third, to markedly change the covering fraction of the \ion{H}{1}
disks, {\it in situ} 
star-formation would have to lower $\approx 50\%$ of the \ion{H}{1}
disks below the DLA criterion.  
At the DLA threshold, star formation
is likely very weak (if not absent) and should not affect this gas.
We conclude, therefore, that {\it in situ} star formation on its
own is insufficient to 
explain the sharp decline in \lox\ and \rhi\ at high $z$.
By similar arguments, one rules out an interpertation where the
majority of \ion{H}{1} gas is simply conveted into molecular gas.
Furthermore, observational biases (e.g.\ dust obscuration, gravitational
lensing) are most important at high \nhi\ values and can be ruled
out as dominant factors.  
Finally, the extragalactic UV background (EUVB) will 
modify the distribution of neutral gas in the outer parts 
of \ion{H}{1} disks but the DLA threshold is sufficiently large that 
variations in the EUVB should play only a minor role for \lox\ and
a negligble one for \rhi\ \citep[e.g.][]{viegas95}.

On the other hand it is possible that the gas in DLAs fuels star
formation \citep{wc06,wolfe08}. 
In that case the Schmidt law would not apply to  DLA gas. Rather, secular
processes could drive the gas to the center 
where it would be consumed by star formation
in compact star-forming regions.
To compute the decrease in {\rhiz}  we fitted analytic
functions to the intrinsic values of {\rhodotz} derived by 
\cite{bif+08} and
computed {\rhiz} by integrating the under the {\rhodotz} curve.
The results are well-matched to the data at $z$=[2.5,4.5] 
(Fig.~\ref{fig:models}b). 
Although we obtained this fit by arbitrarily increasing
the ``initial'' {\rhiz} at $z$=6 by 10 $\%$ above our highest
data point at $z$=4.0, the decrease in {\rhiz} predicted by
this model for {\rhodotz} provides a good match to the data for a wide
range of initial values of {\rhi}.  However, the model breaks down at
$z < 2.4$, where the predicted {\rhiz} falls below the constant
value set by {\rhiz} at $z$=[0,2.4]. But the levelling off of
{\rhi} at $z$ $<$ 2.4, could be explained by delayed
infall of gas from the 
IGM at a rate that balances gas consumption by star formation,
a phenomenon observed in some numerical simulations of galaxy
formation \citep{kkw+05}. Of course, a crucial challenge
to this idea is how to transport gas from the extended regions
comprising DLAs to the center, i.e., by a factor of 15 or
more in radius \citep[cf.][]{wolfe08}, without affecting
the shape of {\fnhi}, but by reducing its normalization
by a factor of two. 

The other extremum is that the sharp decline in covering
fraction and mass density of \ion{H}{1} disks results from `violent'
feedback processes.   By violent, we envision processes that altogether
remove the \ion{H}{1} gas from a galaxy.  These may include AGN activity,
galactic-scale winds, tidal effects, and ram-pressure stripping.
To match the observed evolution in \lox\ and \rhi, one would require
that approximately half of the galaxies exhibiting \ion{H}{1} disks
at $z \approx 4$ have lost their gas by $z=2$.  This implies a dramatic
evolution in the fraction of \ion{H}{1}-rich to \ion{H}{1} poor
galaxies in the 2\,Gyr interval centered at $z\approx 3$.
The resulting galaxies, if unable to accrete new \ion{H}{1} gas
for subsequent star formation, would passively evolve into 
`red and dead' galaxies. 
We speculate, therefore, that $z \sim 3$ marks the formation 
epoch for the formation of stars in the ancestors of modern, 
early-type galaxies. 

The data in Figure~\ref{fig:moments} provides another surprising
result.  The red bands in the figure show $z \approx 0$
estimates of \lox\ and \rhi, 
as inferred from 21cm observations \citep{rws03,zms+05}.
We find that \rhi\ at $z = 2.2$ matches the present-day value.
A brazen, but reasonable, assertion by interpolation
is that \rhi\ has not
evolved over the past 10\,Gyr of our universe.
If confirmed, this result has several important implications.
First, if processes destroy \ion{H}{1} disks below $z=2$,
these must be matched by the formation of new galaxies.
Because the assembly of dark matter halos of galactic-scale masses
($<10^{12} \msol$) is expected to be nearly complete at $z \sim 2$,
we contend that the destruction of \ion{H}{1} disks is also nearly complete. 
Second, the accretion of gas into existing \ion{H}{1} disks
must be balanced by the the consumption of that gas into stars
and/or its removal from the galaxy by feedback processes.
This suggests that \ion{H}{1} disks play a special, but subserviant
role in the formation of stars.  In essence, \ion{H}{1} disks
represent a bias level of gas that is a necessary but 
insufficient condition for star formation.  
In this scenario, the global star formation rate at any given epoch
is driven predominantly by the accretion rate of fresh material
onto existing \ion{H}{1} disks, an inference also drawn from cosmological
simulations of galaxy formation \citep{kkw+05}.
It further suggests that \ion{H}{1} disks have been critically
unstable to star formation over the past 10~billion years and possibly
all of cosmic time.  

Before concluding, we comment briefly on current
estimates of \fnhi\ and \rhi\ at $z \sim 1$ where 
\ion{Mg}{2} absorption has been used to identify DLA candidates
\citep{rao00,rtn06}.  In Figure~\ref{fig:fnhi}b, we present the cumulative
\fnhi\ distribution for DLAs derived in this fashion.
The $z \sim 1$ sample exhibits a higher incidence of DLAs
with $\mnhi > 10^{21} \cm{-2}$ than at $z>2$ or $z \sim 0$.
If taken at face value, the results indicate
that \ion{H}{1} disks
at $z \sim 1$ have a higher cross-section at column densities
$\mnhi \approx 10^{21.6} \cm{-2}$ than $\mnhi \approx 10^{21.3} \cm{-2}$.
It is unlikely that there is a single galaxy in the universe
with this characteristic, much less the integrated population of 
\ion{H}{1} disks.  As such, we conclude that the $z \sim 1$ results
suffer from a statistical fluke or an unfortunate
observational bias.  If we demand
that \fnhi\ at $z \sim 1$ follow the same shape as at $z>2$ and
$z \sim 0$, then we estimate a \rhi\ value consistent with no
evolution over the past 10\,Gyr.

The observations presented here will be supplemented by future
data releases of the SDSS and next generation surveys.   The key
open empirical issues include: 
(i) is there even a mild evolution in the
break column density with redshift? 
(ii) what is the functional form of \fnhi\ beyond the break?
(iii) what are the values of these quantities at $z \approx 6$?
Of greater interest will be to compare the observational constrainst
presented here against theoretical models for the build-up and evolution of
\ion{H}{1} disks over cosmic time.

\acknowledgments

JXP and AMW are supported by NSF grant (AST-0709235).
We are grateful for the tremendous effort put forth by the SDSS team to
produce and release the SDSS survey.  We acknowledge helpful discussions
with M.\ Fumagalli, S. Faber, and J. Primack.  
We thank H.-W. Chen for first suggesting
we draw comparisons with the local universe.  
We also acknowledge the efforts of S. Herbert-Fort who helped build
the algorithms for the DLA survey.


\begin{thebibliography}{30}
\expandafter\ifx\csname natexlab\endcsname\relax\def\natexlab#1{#1}\fi

\bibitem[{{Adelman-McCarthy} {et~al.}(2007){Adelman-McCarthy}, {Ag{\"u}eros},
  {Allam}, {Anderson}, {Anderson}, {Annis}, {Bahcall}, {Bailer-Jones},
  {Baldry}, {Barentine}, {Beers}, {Belokurov}, {Berlind}, {Bernardi},
  {Blanton}, {Bochanski}, {Boroski}, {Bramich}, {Brewington}, {Brinchmann},
  {Brinkmann}, {Brunner}, {Budav{\'a}ri}, {Carey}, {Carliles}, {Carr},
  {Castander}, {Connolly}, {Cool}, {Cunha}, {Csabai}, {Dalcanton}, {Doi},
  {Eisenstein}, {Evans}, {Evans}, {Fan}, {Finkbeiner}, {Friedman}, {Frieman},
  {Fukugita}, {Gillespie}, {Gilmore}, {Glazebrook}, {Gray}, {Grebel}, {Gunn},
  {de Haas}, {Hall}, {Harvanek}, {Hawley}, {Hayes}, {Heckman}, {Hendry},
  {Hennessy}, {Hindsley}, {Hirata}, {Hogan}, {Hogg}, {Holtzman}, {Ichikawa},
  {Ichikawa}, {Ivezi{\'c}}, {Jester}, {Johnston}, {Jorgensen}, {Juri{\'c}},
  {Kauffmann}, {Kent}, {Kleinman}, {Knapp}, {Kniazev}, {Kron}, {Krzesinski},
  {Kuropatkin}, {Lamb}, {Lampeitl}, {Lee}, {Leger}, {Lima}, {Lin}, {Long},
  {Loveday}, {Lupton}, {Mandelbaum}, {Margon}, {Mart{\'{\i}}nez-Delgado},
  {Matsubara}, {McGehee}, {McKay}, {Meiksin}, {Munn}, {Nakajima}, {Nash},
  {Neilsen}, {Newberg}, {Nichol}, {Nieto-Santisteban}, {Nitta}, {Oyaizu},
  {Okamura}, {Ostriker}, {Padmanabhan}, {Park}, {Peoples}, {Pier}, {Pope},
  {Pourbaix}, {Quinn}, {Raddick}, {Re Fiorentin}, {Richards}, {Richmond},
  {Rix}, {Rockosi}, {Schlegel}, {Schneider}, {Scranton}, {Seljak}, {Sheldon},
  {Shimasaku}, {Silvestri}, {Smith}, {Smol{\v c}i{\'c}}, {Snedden}, {Stebbins},
  {Stoughton}, {Strauss}, {SubbaRao}, {Suto}, {Szalay}, {Szapudi}, {Szkody},
  {Tegmark}, {Thakar}, {Tremonti}, {Tucker}, {Uomoto}, {Vanden Berk},
  {Vandenberg}, {Vidrih}, {Vogeley}, {Voges}, {Vogt}, {Weinberg}, {West},
  {White}, {Wilhite}, {Yanny}, {Yocum}, {York}, {Zehavi}, {Zibetti}, \&
  {Zucker}}]{sdssdr5}
{Adelman-McCarthy}, J.~K., {Ag{\"u}eros}, M.~A., {Allam}, S.~S., {Anderson},
  K.~S.~J., {Anderson}, S.~F., {Annis}, J., {Bahcall}, N.~A., {Bailer-Jones},
  C.~A.~L., {Baldry}, I.~K., {Barentine}, J.~C., {Beers}, T.~C., {Belokurov},
  V., {Berlind}, A., {Bernardi}, M., {Blanton}, M.~R., {Bochanski}, J.~J.,
  {Boroski}, W.~N., {Bramich}, D.~M., {Brewington}, H.~J., {Brinchmann}, J.,
  {Brinkmann}, J., {Brunner}, R.~J., {Budav{\'a}ri}, T., {Carey}, L.~N.,
  {Carliles}, S., {Carr}, M.~A., {Castander}, F.~J., {Connolly}, A.~J., {Cool},
  R.~J., {Cunha}, C.~E., {Csabai}, I., {Dalcanton}, J.~J., {Doi}, M.,
  {Eisenstein}, D.~J., {Evans}, M.~L., {Evans}, N.~W., {Fan}, X., {Finkbeiner},
  D.~P., {Friedman}, S.~D., {Frieman}, J.~A., {Fukugita}, M., {Gillespie}, B.,
  {Gilmore}, G., {Glazebrook}, K., {Gray}, J., {Grebel}, E.~K., {Gunn}, J.~E.,
  {de Haas}, E., {Hall}, P.~B., {Harvanek}, M., {Hawley}, S.~L., {Hayes}, J.,
  {Heckman}, T.~M., {Hendry}, J.~S., {Hennessy}, G.~S., {Hindsley}, R.~B.,
  {Hirata}, C.~M., {Hogan}, C.~J., {Hogg}, D.~W., {Holtzman}, J.~A.,
  {Ichikawa}, S.-i., {Ichikawa}, T., {Ivezi{\'c}}, {\v Z}., {Jester}, S.,
  {Johnston}, D.~E., {Jorgensen}, A.~M., {Juri{\'c}}, M., {Kauffmann}, G.,
  {Kent}, S.~M., {Kleinman}, S.~J., {Knapp}, G.~R., {Kniazev}, A.~Y., {Kron},
  R.~G., {Krzesinski}, J., {Kuropatkin}, N., {Lamb}, D.~Q., {Lampeitl}, H.,
  {Lee}, B.~C., {Leger}, R.~F., {Lima}, M., {Lin}, H., {Long}, D.~C.,
  {Loveday}, J., {Lupton}, R.~H., {Mandelbaum}, R., {Margon}, B.,
  {Mart{\'{\i}}nez-Delgado}, D., {Matsubara}, T., {McGehee}, P.~M., {McKay},
  T.~A., {Meiksin}, A., {Munn}, J.~A., {Nakajima}, R., {Nash}, T., {Neilsen},
  Jr., E.~H., {Newberg}, H.~J., {Nichol}, R.~C., {Nieto-Santisteban}, M.,
  {Nitta}, A., {Oyaizu}, H., {Okamura}, S., {Ostriker}, J.~P., {Padmanabhan},
  N., {Park}, C., {Peoples}, J.~J., {Pier}, J.~R., {Pope}, A.~C., {Pourbaix},
  D., {Quinn}, T.~R., {Raddick}, M.~J., {Re Fiorentin}, P., {Richards}, G.~T.,
  {Richmond}, M.~W., {Rix}, H.-W., {Rockosi}, C.~M., {Schlegel}, D.~J.,
  {Schneider}, D.~P., {Scranton}, R., {Seljak}, U., {Sheldon}, E., {Shimasaku},
  K., {Silvestri}, N.~M., {Smith}, J.~A., {Smol{\v c}i{\'c}}, V., {Snedden},
  S.~A., {Stebbins}, A., {Stoughton}, C., {Strauss}, M.~A., {SubbaRao}, M.,
  {Suto}, Y., {Szalay}, A.~S., {Szapudi}, I., {Szkody}, P., {Tegmark}, M.,
  {Thakar}, A.~R., {Tremonti}, C.~A., {Tucker}, D.~L., {Uomoto}, A., {Vanden
  Berk}, D.~E., {Vandenberg}, J., {Vidrih}, S., {Vogeley}, M.~S., {Voges}, W.,
  {Vogt}, N.~P., {Weinberg}, D.~H., {West}, A.~A., {White}, S.~D.~M.,
  {Wilhite}, B., {Yanny}, B., {Yocum}, D.~R., {York}, D.~G., {Zehavi}, I.,
  {Zibetti}, S., \& {Zucker}, D.~B. 2007, \apjs, 172, 634

\bibitem[{{Bouwens} {et~al.}(2008){Bouwens}, {Illingworth}, {Franx}, \&
  {Ford}}]{bif+08}
{Bouwens}, R.~J., {Illingworth}, G.~D., {Franx}, M., \& {Ford}, H. 2008, \apj,
  686, 230

\bibitem[{{Chen} {et~al.}(2005){Chen}, {Prochaska}, {Bloom}, \&
  {Thompson}}]{cpb+05}
{Chen}, H.-W., {Prochaska}, J.~X., {Bloom}, J.~S., \& {Thompson}, I.~B. 2005,
  \apjl, 634, L25

\bibitem[{{Ellison} {et~al.}(2007){Ellison}, {Hennawi}, {Martin}, \&
  {Sommer-Larsen}}]{ehm+07}
{Ellison}, S.~L., {Hennawi}, J.~F., {Martin}, C.~L., \& {Sommer-Larsen}, J.
  2007, \mnras, 378, 801

\bibitem[{{Kennicutt}(1998)}]{kennicutt98}
{Kennicutt}, Jr., R.~C. 1998, \araa, 36, 189

\bibitem[{{Kere{\v s}} {et~al.}(2005){Kere{\v s}}, {Katz}, {Weinberg}, \&
  {Dav{\'e}}}]{kkw+05}
{Kere{\v s}}, D., {Katz}, N., {Weinberg}, D.~H., \& {Dav{\'e}}, R. 2005,
  \mnras, 363, 2

\bibitem[{{Lanzetta} {et~al.}(1995){Lanzetta}, {Wolfe}, \& {Turnshek}}]{lwt95}
{Lanzetta}, K.~M., {Wolfe}, A.~M., \& {Turnshek}, D.~A. 1995, \apj, 440, 435

\bibitem[{{Lin} \& {Pringle}(1987)}]{lp87}
{Lin}, D.~N.~C., \& {Pringle}, J.~E. 1987, \apjl, 320, L87

\bibitem[{{Mo} {et~al.}(1998){Mo}, {Mao}, \& {White}}]{mmw98}
{Mo}, H.~J., {Mao}, S., \& {White}, S.~D.~M. 1998, \mnras, 295, 319

\bibitem[{{Nagamine} {et~al.}(2004){Nagamine}, {Springel}, \&
  {Hernquist}}]{nsh04}
{Nagamine}, K., {Springel}, V., \& {Hernquist}, L. 2004, \mnras, 348, 421

\bibitem[{{Olivier} {et~al.}(1991){Olivier}, {Primack}, \&
  {Blumenthal}}]{obp91}
{Olivier}, S.~S., {Primack}, J.~R., \& {Blumenthal}, G.~R. 1991, \mnras, 252,
  102

\bibitem[{{Pontzen} {et~al.}(2008){Pontzen}, {Governato}, {Pettini}, {Booth},
  {Stinson}, {Wadsley}, {Brooks}, {Quinn}, \& {Haehnelt}}]{pgp+08}
{Pontzen}, A., {Governato}, F., {Pettini}, M., {Booth}, C.~M., {Stinson}, G.,
  {Wadsley}, J., {Brooks}, A., {Quinn}, T., \& {Haehnelt}, M. 2008, \mnras,
  390, 1349

\bibitem[{{Prochaska} {et~al.}(2008){Prochaska}, {Hennawi}, \&
  {Herbert-Fort}}]{phh08}
{Prochaska}, J.~X., {Hennawi}, J.~F., \& {Herbert-Fort}, S. 2008, \apj, 675,
  1002

\bibitem[{{Prochaska} \& {Herbert-Fort}(2004)}]{ph04}
{Prochaska}, J.~X., \& {Herbert-Fort}, S. 2004, \pasp, 116, 622

\bibitem[{{Prochaska} {et~al.}(2005){Prochaska}, {Herbert-Fort}, \&
  {Wolfe}}]{phw05}
{Prochaska}, J.~X., {Herbert-Fort}, S., \& {Wolfe}, A.~M. 2005, \apj, 635, 123

\bibitem[{{Rao} \& {Turnshek}(2000)}]{rao00}
{Rao}, S.~M., \& {Turnshek}, D.~A. 2000, \apjs, 130, 1

\bibitem[{{Rao} {et~al.}(2006){Rao}, {Turnshek}, \& {Nestor}}]{rtn06}
{Rao}, S.~M., {Turnshek}, D.~A., \& {Nestor}, D.~B. 2006, \apj, 636, 610

\bibitem[{{Razoumov} {et~al.}(2006){Razoumov}, {Norman}, {Prochaska}, \&
  {Wolfe}}]{rnp+06}
{Razoumov}, A.~O., {Norman}, M.~L., {Prochaska}, J.~X., \& {Wolfe}, A.~M. 2006,
  \apj, 645, 55

\bibitem[{{Ryan-Weber} {et~al.}(2003){Ryan-Weber}, {Webster}, \&
  {Staveley-Smith}}]{rws03}
{Ryan-Weber}, E.~V., {Webster}, R.~L., \& {Staveley-Smith}, L. 2003, \mnras,
  343, 1195

\bibitem[{{Schaye}(2001)}]{schaye01_H2}
{Schaye}, J. 2001, \apjl, 562, L95

\bibitem[{{Schmidt}(1959)}]{schmidt59}
{Schmidt}, M. 1959, \apj, 129, 243

\bibitem[{{Viegas}(1995)}]{viegas95}
{Viegas}, S.~M. 1995, \mnras, 276, 268

\bibitem[{{Walter} {et~al.}(2008){Walter}, {Brinks}, {de Blok}, {Bigiel},
  {Kennicutt}, {Jr.}, {Thornley}, \& {Leroy}}]{things08}
{Walter}, F., {Brinks}, E., {de Blok}, W.~J.~G., {Bigiel}, F., {Kennicutt},
  R.~C., {Jr.}, {Thornley}, M.~D., \& {Leroy}, A.~K. 2008, ArXiv e-prints

\bibitem[{{Wolfe} \& {Chen}(2006)}]{wc06}
{Wolfe}, A.~M., \& {Chen}, H.-W. 2006, \apj, 652, 981

\bibitem[{{Wolfe} {et~al.}(2005){Wolfe}, {Gawiser}, \& {Prochaska}}]{wgp05}
{Wolfe}, A.~M., {Gawiser}, E., \& {Prochaska}, J.~X. 2005, \araa, 43, 861

\bibitem[{{Wolfe} {et~al.}(2008){Wolfe}, {Jorgenson}, {Robishaw}, {Heiles}, \&
  {Prochaska}}]{wolfe08}
{Wolfe}, A.~M., {Jorgenson}, R.~A., {Robishaw}, T., {Heiles}, C., \&
  {Prochaska}, J.~X. 2008, \nat, 455, 638

\bibitem[{{Wolfe} {et~al.}(1986){Wolfe}, {Turnshek}, {Smith}, \&
  {Cohen}}]{wolfe86}
{Wolfe}, A.~M., {Turnshek}, D.~A., {Smith}, H.~E., \& {Cohen}, R.~D. 1986,
  \apjs, 61, 249

\bibitem[{{Zwaan} {et~al.}(2005{\natexlab{a}}){Zwaan}, {Meyer},
  {Staveley-Smith}, \& {Webster}}]{zms+05}
{Zwaan}, M.~A., {Meyer}, M.~J., {Staveley-Smith}, L., \& {Webster}, R.~L.
  2005{\natexlab{a}}, \mnras, 359, L30

\bibitem[{{Zwaan} \& {Prochaska}(2006)}]{zp06}
{Zwaan}, M.~A., \& {Prochaska}, J.~X. 2006, \apj, 643, 675

\bibitem[{{Zwaan} {et~al.}(2005{\natexlab{b}}){Zwaan}, {van der Hulst},
  {Briggs}, {Verheijen}, \& {Ryan-Weber}}]{zvb+05}
{Zwaan}, M.~A., {van der Hulst}, J.~M., {Briggs}, F.~H., {Verheijen}, M.~A.~W.,
  \& {Ryan-Weber}, E.~V. 2005{\natexlab{b}}, \mnras, 364, 1467

\end{thebibliography}

\clearpage



\end{document}